Bruna Contro[1,*], Rob Wittenmyer[1,2,3], Jonti Horner[2,3],
Jonathan P. Marshall[1,3]

# The dynamical structure of HR 8799's inner debris disk

[1]School of Physics, UNSW Australia, Sydney, New South Wales 2052, Australia
[2]Computational Engineering and Science Research Centre, University of Southern Queensland, Toowoomba, Queensland 4350, Australia
[3]Australian Centre for Astrobiology, UNSW Australia, Sydney, New South Wales 2052, Australia

[*]contro.bc@gmail.com

**Abstract**

The HR 8799 system, with its four giant planets and two debris belts, has an architecture closely mirroring that of our Solar system where the inner, warm asteroid belt and outer, cool Edgeworth-Kuiper belt bracket the giant planets. As such, it is a valuable laboratory for examining exoplanetary dynamics and debris disk-exoplanet interactions. Whilst the outer debris belt of HR 8799 has been well resolved by previous observations, the spatial extent of the inner disk remains unknown. This leaves a significant question mark over both the location of the planetesimals responsible for producing the belt's visible dust and the physical properties of those grains. We have performed the most extensive simulations to date of the inner, unresolved debris belt around HR 8799, using UNSW Australia's *Katana* supercomputing facility to follow the dynamical evolution of a model inner disk comprising 300,298 particles for a period of 60 million years. These simulations have enabled the characterisation of the extent and structure of the inner disk in detail, and will in future allow us to provide a first estimate of the small-body impact rate and water delivery prospects for possible (as-yet undetected) terrestrial planet(s) in the inner system.

**Introduction**

In recent years, a vast number of planets have been detected orbiting other stars. Since the first discoveries, which featured hot, massive planets, orbiting close to their host stars (e.g. 51 Pegasi; Mayor & Queloz, 1995), we have gradually moved towards the detection of the first true Solar system analogues – planetary systems that closely resemble our own. The *Kepler* spacecraft has played a critical role in detecting the smallest planets found to date[1] (e.g. Kepler 37b, which is significantly smaller than the planet Mercury; Barclay *et al.*, 2013), and so could, in theory, detect truly Earth-like worlds. However, to date, no such discovery has been forthcoming – owing to the spacecraft's loss of fine pointing, in 2013, following the failure of two reaction wheels. In the coming years, it is likely that the discovery of such planets (exoEarths) will be announced using data taken by *Kepler*, allowing astronomers, for the first time, to draw conclusions on how common (or unusual) are planets like our own.

Whilst transiting exoplanet observation programs such as *Kepler* continue to push the boundaries of our knowledge to ever smaller planets orbiting close to their host stars, direct imaging and radial velocity programs have pushed outwards, finding massive planets at ever greater distances from their host stars. Two decades after the discovery of 51 Pegasi b, we are finally finding giant planets moving on orbits comparable to that of Jupiter in our own Solar system, and so stand on the cusp of detecting the first planetary systems that truly resemble our own (e.g. Wittenmyer et al., 2011; 2013; 2014).

Taken together, these two distinct streams of exoplanetary research are a vital first step in our attempts to search for evidence of life beyond the Solar system. Though it is certainly feasible that life could develop and thrive in planetary systems far different to our own, in our first attempts to search for life it behoves us to search in those locations that seem most likely to yield a positive result. In other words, the first searches for life beyond the Solar system will search for the signs of life like that found on the Earth, and will therefore target the planets deemed most promising as hosts of such life. A wide variety of factors have been proposed that could render a given exoEarth more (or less) habitable than others we find (e.g. Horner & Jones, 2010) – but it is clear that a key motivator to search a given planetary system will be the degree to which it resembles our own – i.e., the degree to which it is a *Solar system analogue*.

Despite the rapid progress astronomers have made in the discovery of exoplanets, technological limitations have meant that, to date, very few extrasolar planetary systems have been found that closely parallel the architecture of our Solar system. One such example is HR 8799 (HD 218396). Whilst, to date, no Earth-like planets have been discovered in that system, HR 8799 hosts at least four giant planets (e.g. Marois et al. 2010) and a circumstellar disc (Rhee *et al.* 2007) comprised of two debris belts (e.g. Matthews *et al.* 2014).

Whilst the great majority of known exoplanets were discovered using either the radial velocity or transit techniques, the planets orbiting HR 8799 were instead discovered through direct imaging of the system[2]. The first three planets found in the

---

[1] *Kepler* has discovered a total of 989 confirmed planets, as of 7th October 2014, with their website (http://kepler.nasa.gov) detailing a further 4234 candidate planets awaiting validation.
[2] A detailed description of the various techniques used to detect exoplanets is beyond the scope of this conference paper, and we direct the interested reader to

system (HR 8799 b, c and d) were discovered in 2008, as a result of adaptive optics observations on the Keck and Gemini telescopes (Marois *et al.*, 2008). The discovery of the fourth and innermost planet found to date, HR 8799 e, was announced from follow-up observations by the same group in 2010 (Marois *et al.*, 2010). A recent dynamical study (Gozdziewski & Migaszewski, 2014) suggests that, assuming the planets in the HR 8799 underwent rapid migration as they formed, they are most likely currently trapped in a double Laplace resonance. In this configuration, the planets have orbital period ratios of 1:2:4:8. A schematic showing the motion of the four planets in the system is shown in Figure 1. In this resonant architecture, the orbits of the planets can remain dynamically stable on long timescales. In that figure, the pale blue shaded area, interior to the orbit of HR 8799 e, represents the approximate radius of the "last stable orbit" determined in that work – and likely matches the location of the inner of the two debris disks detected in the system, the focus of this work.

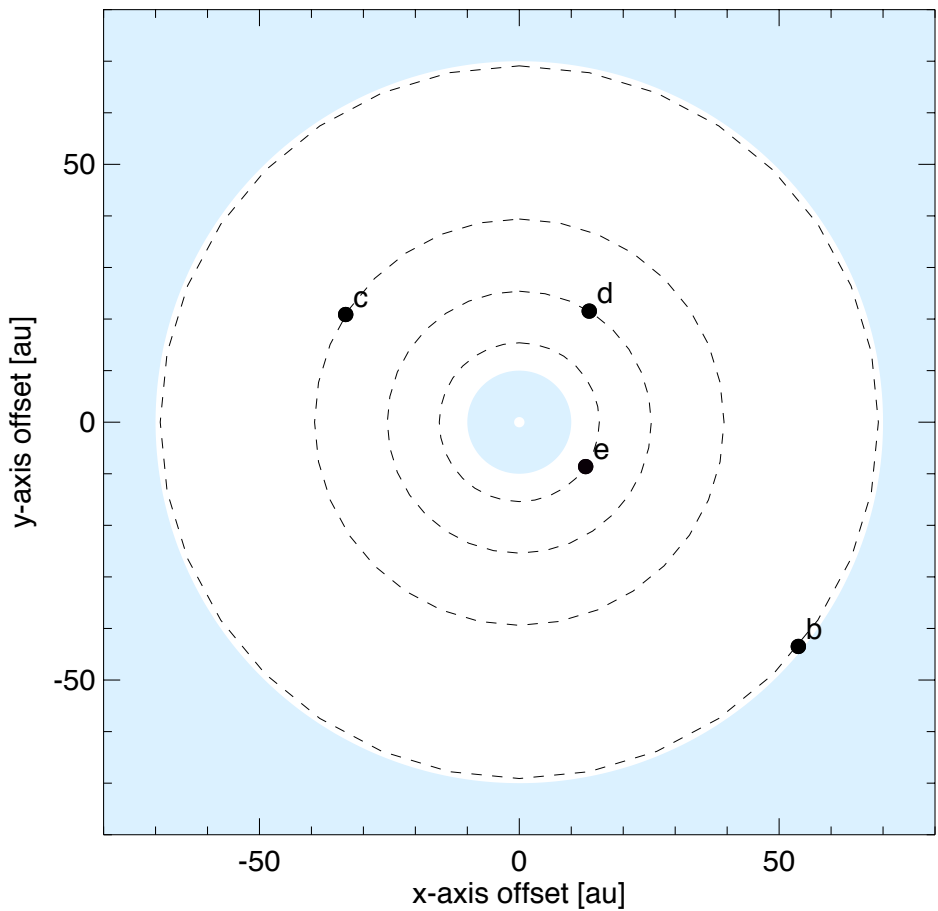

**Figure 1.** This is a schematic plot of the HR 8799 system. The four planets are at their correct orbital separations from the star, and location around the star, but are on circular rather than eccentric orbits. The light blue regions are those parts of the system where circumstellar

---

the Exoplanet Handbook (Perryman, 2012, 2014) for a detailed and thorough review of the field.

debris must be located in order to explain the observed infra-red excess described in Matthews et al., 2014.

HR 8799 is a relatively young A-type star, located approximately 40 pc from the Earth (van Leeuwen 2007). Its age remains relatively poorly determined, although its youth is well accepted. Reidemeister *et al.* (2009) suggest that the star is no more than 50 Myr old. That result is supported by the work of Zuckerman *et al.* (2011), who suggest an age of 30 Myr for the star, based on its likely membership of the Columbia Association, a group of young stars spread across the sky that share a common movement and common origin. To be conservative, Marois *et al.* (2010) consider two different ages for the HR 8799 planetary system – 30 Myr (based on the Columbia Association), and 60 Myr (following their earlier work, Marois *et al.*, 2008). The age of the system is critically important in the determination of the mass of the planets therein – older planets would have cooled more since their formation, and therefore be fainter at the current epoch than younger planets of the same mass. See Table 1 for a summary of the stellar parameters used in this work.

**Table 1**
Stellar Parameters for HR 8799.

| Parameter | Value | Reference |
|---|---|---|
| Age (Myr) | 30 | Zuckerman *et al.* (2011) |
|  | 60 | Marois *et al.* (2008) |
| Teff (K) | 7193 ± 87 | Baines *et al.* (2012) |
| L ($L_\odot$) | 5.05 ± 0.29 | Baines *et al.* (2012) |

In addition to the four planets detected by Marois *et al.* (2010), the HR 8799 system has also been found to host two debris belts. The outer belt has been spatially resolved (Rhee *et al.* 2007; Reidemeister *et al.* 2009; Matthews *et al.* 2014), with a belt structure extending from 100 (±10 au) to 310 au, and a halo of small grains out to > 1000 au beyond (Matthews et al. 2014), In contrast, the inner belt is, to date, poorly understood (Matthews *et al.* 2014). Our current best estimate of its location is simply that it must lie somewhere between 1 and 10 au, based on a black body fit to the observed dust temperature, which is a good approximation for A stars (Booth *et al.* 2013; Pawellek *et al.* 2014). This information, coupled with the orbits and masses of the giant planets beyond the disk, gives us enough information to attempt to constrain its extent and interior structure using *n*-body simulations.

Our interest in the inner reaches of the HR 8799 system is driven by its apparent similarity to our own Solar system. Whilst it is almost certainly too young a system to be considered a viable host for complex life, it remains a fascinating test case to compare to our own system. Since HR 8799 is significantly hotter and more luminous than the Sun, the "Habitable Zone" (HZ; the region around the star in which an Earth like planet might reasonably be expected to be able to host liquid water on its surface, e.g. Kasting, 1993) would lie at a greater astrocentric distance. However, that region would be well inside the orbit of HR 8799 e, and may well be interior to the inner disk (just as, in our Solar system, the HZ falls well interior to the inner edge of the asteroid belt). Simulations of the sculpting of the inner disk can then tell us about its likely structure, but also give some indication of the potential impact rates that would be experienced by any exoEarths orbiting in the system (e.g. Horner & Jones, 2008), and might even shed light on the potential delivery of volatiles to such planets

(e.g. Owen & Bar-Nun, 1995, Horner et al., 2009). The most probable place to find a potentially habitable terrestrial planet would then be where the impact rates are not sufficiently high that they would disrupt the development of life, but high enough to allow the delivery of water to planets that might otherwise have formed arid.

**Material and Methods**

We perform dynamical simulations of the inner debris belt around HR 8799 using the numerical integrator package MERCURY (Chambers 1999), using UNSW's Katana supercomputing facility. These preliminary integrations featured a disk that was initially composed of 300,298 massless test particles. Each of the six orbital elements for every test particle was randomly allocated, to lie within a set range. The semi-major axes were randomly chosen to lie between 1 and 10 au (with the inner and outer edges located in accordance with the thermal modelling of the observed disk described in Su *et al.* 2009; Matthews *et al.* 2014). Once the semi-major axis had been determined, the eccentricity of the particle was randomly allocated to lie between 0.1 and 1.0. Each particle was then given a randomly determined orbital inclination between 0 and 5 degrees. The rotational orbital elements for each particle were each separately randomly assigned values between 0 and 360 degrees. The end result of this process was a dynamically excited disk of debris that filled the inner reaches of the HR 8799 system.

Once the suite of test particles had been created, its dynamical evolution was then followed for a period of 60 Myr (to ensure compatibility with both the younger and older ages used in Marois *et al.* 2008 and 2010), under the gravitational influence of the four giant planets known in the system, with an integration time-step of 7 days. Each individual test particle was followed until it collided with the central body, impacted on one of the giant planets, or was ejected from the system (upon reaching a barycentric distance of 1000 AU). The orbital elements of all surviving test particles were then output at 6 Myr intervals, allowing us to trace the evolution of the disk's extent and architecture over the lifetime of the system.

For this study, we followed the best-fit parameters for HR 8799 b, c, d and e, and a stellar mass of $1.56 M_\odot$ described by Gozdziewski and Migaszewski (2014), as shown in Table 2. In addition, the location of the classical HZ was estimated to be 1.974 (inner edge) to 3.407 au (outer edge), following the prescriptions given in Selsis et al. (2007) and Kopparapu et al. (2014), using the stellar parameters detailed in Table 1.

**Table 2**
Orbital elements for the planetary system around HR 8799, as used in our integrations. These elements are those of the double-Laplace resonant architecture proposed in Table 1 of Gozdziewski & Migaszewski (2014). Here, *m* is the mass of the planet, in Jupiter masses, *a* is the orbital semi-major axis, in au, and *e* is the eccentricity of the orbit. *i* is the inclination of the orbit to the plane of the sky, and $\Omega$, $\omega$ and *M* are the longitude of the ascending node, the longitude of pericentre, and the mean anomaly at the epoch 1998.83, respectively.

|  | *m* [$m_{jup}$] | *a* [au] | e | *i* [deg] | $\Omega$ [deg] | $\omega$ [deg] | *M* [deg] |
|---|---|---|---|---|---|---|---|
| HR 8799 e | 9±2 | 15.4±0.2 | 0.13±0.03 | 25±3 | 64±3 | 176±3 | 326±5 |
| HR 8799 d | 9±2 | 25.4±0.3 | 0.12±0.02 |  |  | 91±3 | 58±3 |

| | | | | | | |
|---|---|---|---|---|---|---|
| HR 8799 c | 9±2 | 39.4±0.3 | 0.05±0.02 | | 151±6 | 148±6 |
| HR 8799 b | 7±2 | 69.1±0.2 | 0.020±0.003 | | 95±10 | 321±10 |

*No humans or animals were used as experimental subjects in this research.*

## Results

The instantaneous semi-major axes and eccentricities of the test particles in the inner debris disk of HR 8799 are shown as a function of time in Figure 2. Panel one shows the initial orbits of the 300,298 test particles, whilst the other panels show the elements of those particles surviving at each of the time-steps in question (6 Myr intervals between 6 and 54 Myr). Finally, Figure 3 shows in more detail the final distribution of the surviving test particles after the full 60 Myr have elapsed.

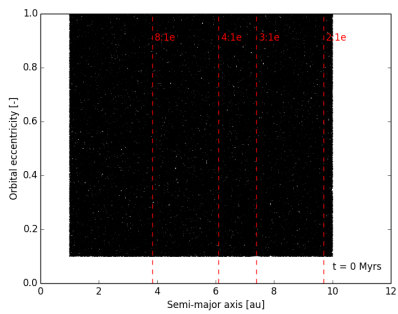
**Panel 1:** 0 Myr.

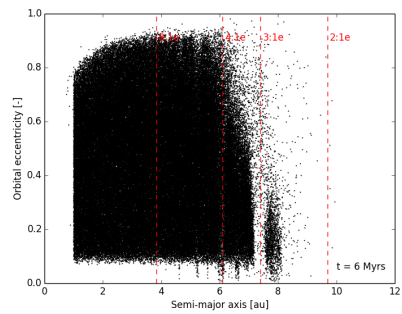
**Panel 2:** 6 Myr.

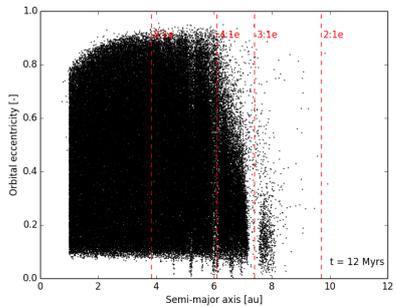
**Panel 3:** 12 Myr.

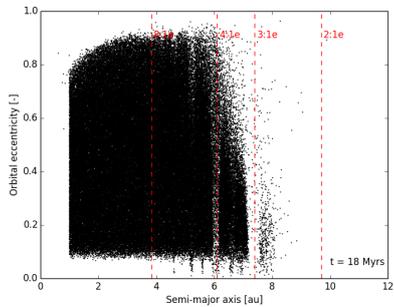
**Panel 4:** 18 Myr.

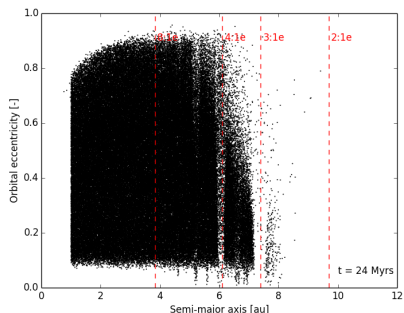
**Panel 5:** 24 Myr.

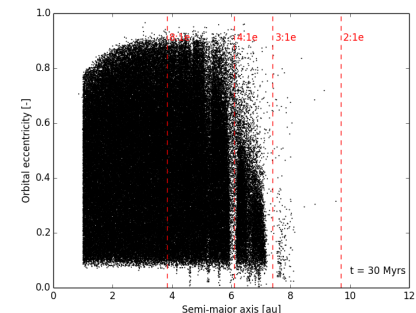
**Panel 6:** 30 Myr.

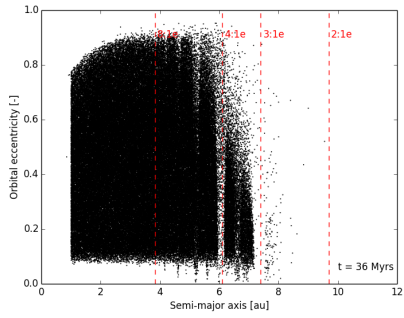
**Panel 7:** 36 Myr.

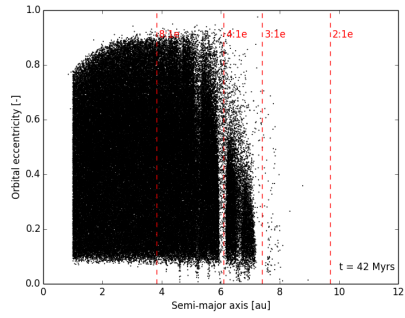
**Panel 8:** 42 Myr.

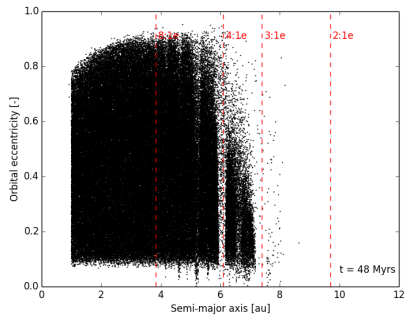
**Panel 9:** 48 Myr.

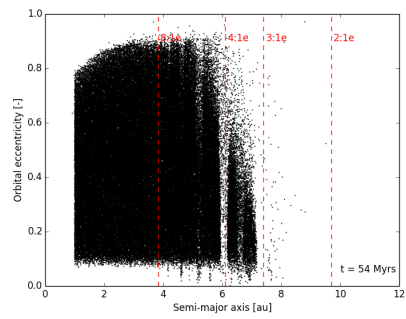
**Panel 10:** 54 Myr.

**Figure 2.** The distribution of surviving test particles orbiting HR 8799 in our simulations, as a function of time, from t = 0 to t = 54 Myr, in semi-major axis-eccentricity space. Note the rapid initial dispersal of debris beyond ~8 au, followed by a more sedentary sculpting of the disk over the remaining duration of the integration.

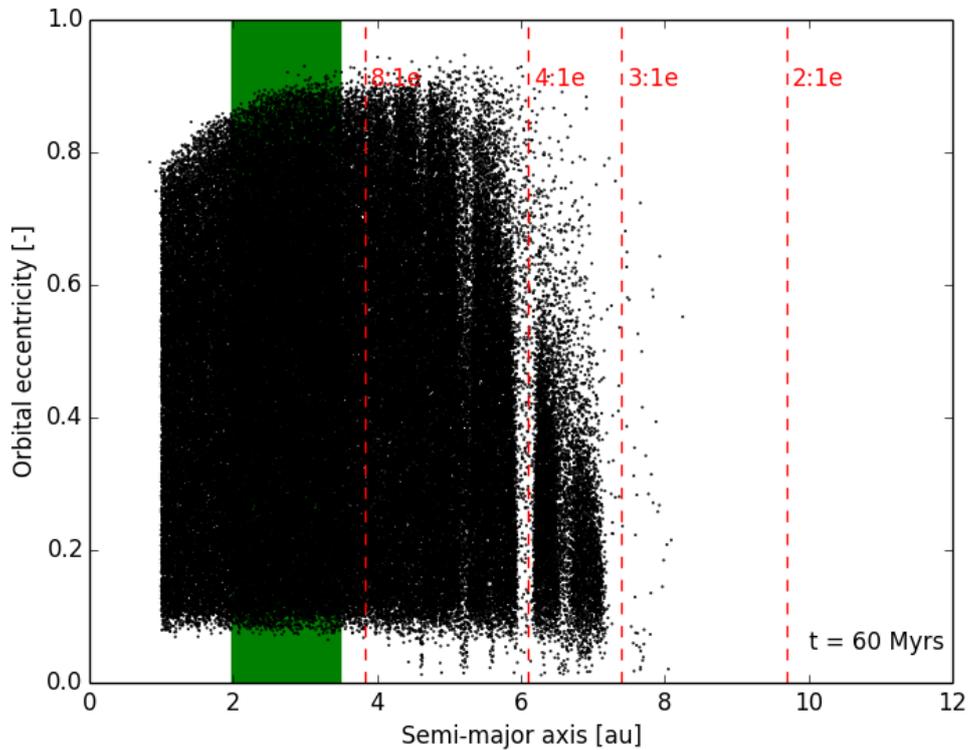

**Figure 3.** The final distribution of test particles orbiting HR 8799 in semi-major axis – eccentricity space after 60 Myr. The sculpting influence of mean-motion resonances between debris and planets can be clearly seen both in the gaps introduced to the disk (e.g. just outside 6 AU), and in those locations where test particles have been driven to stable orbits with lower eccentricities (the downward "spikes" visible throughout the plot).

**Discussion**

Over the course of the 60 Myr integrations described in the previous sections, the disk of debris orbiting HR 8799 undergoes drastic sculpting. The most readily apparent result of this sculpting is the rapid and complete clearing of debris from the outer region of the disk. Simply put, debris in this region is dynamically untenable – within the first few million years of our integrations, the innermost planet clears this region of debris – just as Jupiter has swept the region between the outer edge of our asteroid belt and its orbit clean. Aside from a few stray objects, the disk at the end of our integrations is sharply truncated at an astrocentric distance of just ~7 au. This places a new constraint on the maximum radial extent of the disk that is significantly stricter than that which could be achieved purely on the basis of the thermal modelling described in Booth *et al.* (2013), bringing the putative outer edge of the disk inward from 10 au to ~7 au.

Looking through the results, the extremely rapid dispersal of distant and unstable objects is readily apparent, with significant changes to the structure of the disk visible after just 6 Myr (Panel 2 on Figure 2). In this first 6 Myr period, objects exterior to ~8 au are removed, shaping the disk's outer edge, whilst resonant perturbations act to sculpt the disk at smaller astrocentric radii. However, after 6 Myr, the sculpting slows, which can be seen by comparing the various panels in Figure 2 as one moves towards the final result. This suggests that the initial clearance of the outermost and unstable debris is a very rapid and chaotic process, as evidenced by the rapid initial decay of the test particle population.

At the final time-step (shown in detail in Figure 3), the structure of the disk has become clearly defined. A sharp outer edge can be seen at approximately 7 au revealing that, in general, debris cannot survive beyond this distance without being rapidly removed by the influence of the outer planets. However, a small population of objects remains at around 7.5 au. These objects appear to be trapped in 3:1 mean motion resonance with HR 8799 e (a resonance centred at ~7.4 au).

Within the main disk, several additional resonant features are evident. The most readily apparent of these are the gap at approximately 6.1 au (the location of the 4:1 mean motion resonance with HR 8799 e), and low eccentricity objects between 4.5 and 5.5 au in narrow strips at the location of other, higher order resonances. In addition, a clear sculpting can be seen at high eccentricities at the inner edge of the disk. These highly eccentric objects move on orbits with such small periastra and orbital periods that the 7-day time-step used for these integrations is insufficient. This feature is therefore a computational effect, rather than the result of a real dynamical process.

Our results show that the giant planets not only affect the extent of the debris, but also its fine structure – just as the structure of the asteroid belt is heavily sculpted by our own giant planets. Complementary studies are being done to investigate both the mean-motion resonances and the impact rates and water delivery to the HZ. Further work will yield a better understanding of the dynamical structure of the inner debris belt, and offer clues as to the possibility and viability of a fifth, terrestrial planet in the HZ.

**Acknowledgements**

This work is supported by CAPES Foundation, Ministry of Education of Brazil, Brasilia – DF, Zip Code 70.040-020 and School of Physics - UNSW Australia, and made use of UNSW Australia's *Katana* supercomputing cluster.

**Conflict of Interest**

The authors do not work for, consult to, own shares in or receive funding from any entity that would benefit from this article. As such, they have no conflicts of interest to declare.